\documentclass[]{interact}
\usepackage{color,soul}
\usepackage{epstopdf}% To incorporate .eps illustrations using PDFLaTeX, etc.
\usepackage[caption=false]{subfig}% Support for small, `sub' figures and tables
\usepackage[T1]{fontenc}
\usepackage{textcomp}

\usepackage[numbers,sort&compress]{natbib}% Citation support using natbib.sty
\bibpunct[, ]{[}{]}{,}{n}{,}{,}% Citation support using natbib.sty
\makeatletter% @ becomes a letter
\def\NAT@def@citea{\def\@citea{\NAT@separator}}% Suppress spaces between citations using natbib.sty
\makeatother% @ becomes a symbol again

\theoremstyle{plain}% Theorem-like structures provided by amsthm.sty

\theoremstyle{definition}

\theoremstyle{remark}

\begin{document}

\articletype{Article}% Specify the article type or omit as appropriate

\title{A generalized Young's equation to bridge a gap between the experimentally measured and the theoretically calculated line tensions}

\author{
\name{Masao Iwamatsu\textsuperscript{ab}\thanks{CONTACT M. Iwamatsu. Email: iwamatm@tcu.ac.jp}}
\affil{\textsuperscript{a}Department of Physics, Tokyo City University, Setagaya-ku, Tokyo 158-8557, Japan; \textsuperscript{b}Department of Physics, Tokyo Metropolitan University, Hachioji, Tokyo 192-0397, Japan}
}

\maketitle

\begin{abstract}
A generalized Young's equation, which takes into account two corrections to the line tension by the curvature dependence of the liquid-vapor surface tension and by the contact angle dependence of the intrinsic line tension, is derived from the thermodynamic free-energy minimization.  The correction from the curvature dependence can be qualitatively estimated using Tolman's formula.  The correction from the contact angle dependence can be estimated for nanometer-scale droplets for which the analytical formula for the intrinsic line tension determined from the van der Waals interaction is available.  The two corrections to the apparent line tension of this van der Waals nano-droplets are as small as nN, and lead to either a positive or a negative apparent line tension.   The gravitational line tension for millimeter-scale droplets by the gravitational acceleration is also considered.  The gravitational line tension is of the order of $\mu$N so that the correction from the curvature dependence can be neglected.  Yet, the contact angle dependence is so large that the apparent line tension becomes always negative though the intrinsic line tension without the correction is always positive.  These two examples demonstrate clear distinction between the theoretical calculated intrinsic line tension and the experimentally determined apparent line tension which includes these two corrections.  Naive comparison of the experimentally determined and the theoretically calculated line tension is not always possible.
\end{abstract}

\begin{keywords}
Line tension; contact angle; Young's formula; Tolman's formula 
\end{keywords}

\section{\label{sec:sce1}Introduction}
\begin{figure}[htbp]
%Fig.1
\begin{center}
\includegraphics[width=0.85\linewidth]{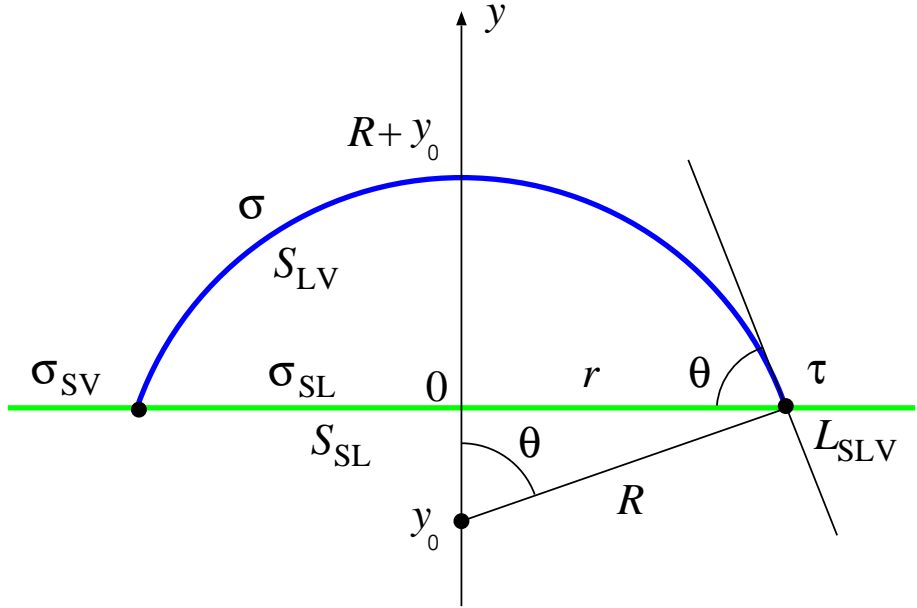}
\caption{
A spherical cap-shaped droplet with a radius $R$ on a flat substrate.   The center of the spherical meniscus locates at $y_{0}=-R\cos\theta$, where $\theta$ is the contact angle.  The radius of three-phase-contact line is given by $r=R\sin\theta$.  The Helmholtz free energy of the droplet consists of the liquid-vapor surface tension $\sigma$ multiplied by the surface area $S_{\rm LV}$, the difference of the solid-liquid surface tension $\sigma_{\rm SL}$ and the solid-vapor surface tension $\sigma_{\rm LV}$ multiplied by the surface area $S_{\rm SL}$, and the line tension $\tau$ multiplied by the three-phase contact line length $L_{\rm SLV}$. 
}
\label{fig:L1}
\end{center}
\end{figure}

The sign and the magnitude of experimentally measured line tension are still the issue of debate~\cite{Law2017}, although the concept of line tension has been established more than a century ago in seventieth of nineteenth century by Gibbs~\cite{Gibbs1878,Rowlinson1982,deGennes2004}.  The well-known starting point for the experimental study of line tension $\tau$ is the size-dependent contact angle $\theta$ of a droplet place on a flat and ideally smooth solid substrate (Fig.~\ref{fig:L1}) through the so-called modified Young's equation~\cite{Law2017,Boruvka1977}
\begin{equation}
\cos\theta=\cos\theta_{\rm Y}-\frac{\tau}{\sigma r},
\label{eq:T1}
\end{equation}
where $r$ is the radius of (circular) contact line, $\sigma$ is the liquid-vapor surface tension, and $\theta_{\rm Y}$ is the Young's contact angle defined through the Young's equation~\cite{Young1805,Rowlinson1982}
\begin{equation}
\cos\theta_{\rm Y}=\frac{\sigma_{\rm SL}-\sigma_{\rm SV}}{\sigma}
\label{eq:T2}
\end{equation}
with $\sigma_{\rm SL}$ as the solid-liquid surface tension, and $\sigma_{\rm SV}$ as the solid-vapor surface tension~\cite{Rowlinson1982,deGennes2004,Law2017}.  If the droplet is cylindrical with a straight contact line ($r\rightarrow\infty$), which cannot be realized experimentally but easily realized by computer simulations~\cite{Maheshwari2016,Kanduc2017}, the last term in Eq.~(\ref{eq:T1}) is missing and the contact angle $\theta$ is given simply by the Young's contact angle $\theta_{\rm Y}$ in Eq.~(\ref{eq:T2}) and does not depend on the size of the droplet.  On the other hand, if the droplet is a spherical cap with radius $R$ shown in Fig.~\ref{fig:L1}, the radius $r$ of the contact line is given by $r=R\sin\theta$ and the contact angle $\theta$ depends on the size $R$ of the droplet.

By plotting the cosine of the contact angle $\theta$ ($\cos\theta$) versus curvature $1/r$ by changing the base radius ($r$) of the droplet, one can deduce the sign and the magnitude of the line tension $\tau$ from Eq.~(\ref{eq:T1}) if the line tension $\tau$ does not depend on the size $r$ or the contact angle $\theta$.  The line tension $\tau$ has been determined~\cite{Law2017} from millimeter-scale droplets by optical microscopes down to nanometer-scale droplets by atomic-force microscopes.  The experimentally determined line tensions can be either positive or negative, and differ several order of magnitude~\cite{Law2017,Drelich1996}: The millimeter-scale droplet has a line tension as large as $10^{-5}$N while the nanometer-scale droplet has a line tension as small as $10^{-10}-10^{-12}$N~\cite{Law2017,Drelich1996,Pompe2000,Wang2001,Checco2003,Kameda2008,Berg2010}.  Therefore, the line tension depends on the size scale of droplets~\cite{David2007,Heim2013}.  Furthermore, the deviation from the linearity $\cos\theta\propto 1/r$ of Eq.~(\ref{eq:T1}) is frequently observed~\cite{Checco2003,Heim2013}, which is attributed partly to the fact that the line tension $\tau$ depends on the contact angle $\theta$~\cite{Heim2013}.  Even the reversal of the sign of line tension is observed by changing the temperature~\cite{Law2017,Wang2001} or the size of droplets~\cite{Kameda2008}.  This modified Young equation (\ref{eq:T1}) is routinely used to deduce the magnitude of line tension from the computer simulation as well~\cite{Weijs2011,Liu2013,Maheshwari2016,Kanduc2017}.  Therefore, the experimentally determined line tension, which we call {\it apparent} line tension, is defined through the modified Young's equation in Eq.~(\ref{eq:T1}).  Since Eq.~(\ref{eq:T1}) represents the force balance parallel to the surface at the three-phase contact line~\cite{Iwamatsu2015}, the apparent line tension is defined through the {\it force} rather than through the {\it energy}.

Theoretically, the line tension is defined as the excess contribution to the free energy associated with the three-phase-contact line scales with its length~\cite{Gibbs1878,Navascues1981,Widom1995,Schimmele2007,Schimmele2009}.  The line contribution of free energy per unit length of the contact line is called line tension $\tau$, which we call {\it intrinsic} line tension. By decomposing the free energy of a droplet on a solid substrate into volume, interfacial, and line contributions, we can theoretically determine the intrinsic line tension.  Since the liquid-vapor surface is curved and interfacial zone is diffuse, there is a conceptual problem of dividing surface~\cite{Kondo1955,Rowlinson1982,Rusanov2004,Schimmele2007,Wang2014}. Therefore, instead of directly analyzing a cap-shaped droplet with a finite base radius and a spherical meniscus, a liquid wedge with a flat meniscus, which corresponds to the limit of infinite base radius, is usually considered.  So far, the most of microscopic calculations of intrinsic line tension assume a sharp interface and a straight ($r\rightarrow \infty$) three-phase-contact line and a planar liquid-vapor interfaces~\cite{Indekeu1992,Getta1998,Pompe2002,Herminghaus2006,Matsubara2018}. 

Since the vicinity of the three-phase-contact line is the region where the multiple scale length from macroscopic millimeter to atomic sub-nanometer scale meets~\cite{Law2017,Chen2014}, different physical phenomena at different length scales contribute to the line tension~\cite{Law2017}.  At sub-nanometer atomic scale, a direct atomic interaction between different species of atoms contributes to the atomic line tension $\tau_{\rm atom}$, whose magnitude is~\cite{Law2017} $\left|\tau_{\rm atom}\right|\sim 1$nN.  At nanometer scale, the direct surface force such as the van der Waals interaction usually modeled by the surface potential or the disjoining potential~\cite{Indekeu1992,Getta1998,Pompe2002} plays dominant role, which leads to the van der Waals line tension~\cite{Law2017} $\left|\tau_{\rm vdW}\right|\sim 1-100$pN.  At a larger millimeter scale of the order of capillary length $\kappa^{-1}\sim 1$mm, the gravitational line tension~\cite{deGennes2004,Herminghaus2006,Law2017} $\tau_{\rm grav}\sim 1-10(>0)\mu{\rm N}$ is dominant.   Observed line tension would be the sum of those three contributions~\cite{Law2017}:
\begin{equation}
\tau=\tau_{\rm atom}+\tau_{\rm vdW}+\tau_{\rm grav}.
\label{eq:T3}
\end{equation}
For nanometer-scale droplet, of course, sum must be up to the first two terms $\tau_{\rm atom}$ and $\tau_{\rm vdW}$.

Furthermore, the experimentally determined apparent line tension $\tau_{\rm app}$ from Eq.~(\ref{eq:T1}) always contains combinations of all the different $1/r$ corrections to the free energy.  For example, the liquid-vapor surface tension $\sigma$ is not a constant but depends on the curvature of the spherical liquid-vapor interface through the so-called Tolman's length~\cite{Tolman1949,Iwamatsu1994,vanGeesen1998,Liu2013,Kanduc2017}.   However, the recent molecular simulations demonstrate that both the curvature dependence of the liquid-vapor surface tension represented by Toman's length~\cite{Liu2013,Kanduc2017} and the contact-angle dependence of the intrinsic line tension~\cite{Heim2013,Kanduc2017} cannot be neglected.  In fact, Eq.~(\ref{eq:T1}) indicates that the line tension should depend on the contact angle $\theta$ or the radius $r=R\sin\theta$ of the three-phase-contact line of a spherical-cap shaped droplet as the last term of Eq.~(\ref{eq:T1}) diverges~\cite{Iwamatsu2017a} when $\theta\rightarrow 0$ or $r\rightarrow 0$ on a spherical substrate.  

In this paper, we will derive a generalized Young's equation which takes into account the effect of the curvature dependence of liquid-vapor surface tension and the contact-angle dependence of intrinsic line tension.  Then, we will consider the van der Waals and the gravitational line tensions as two examples.

\section{\label{sec:sce2}A generalized Yount's equation}

\begin{figure}[htbp]
%Fig.2
\begin{center}
\includegraphics[width=0.95\linewidth]{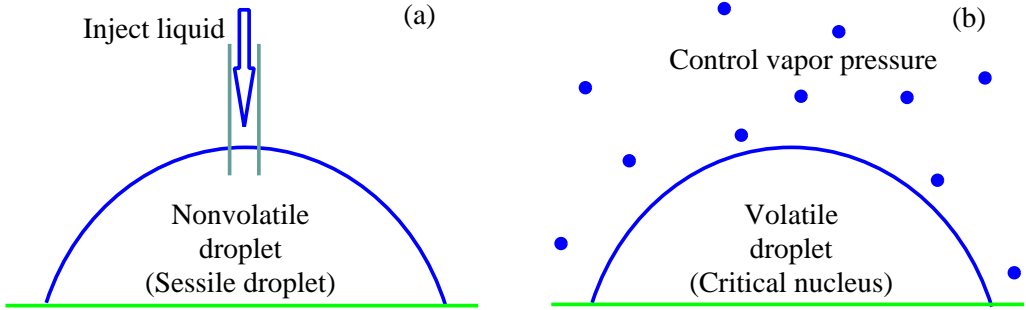}
\caption{
(a) A nonvolatile droplet (sessile droplet) on a flat and smooth substrate.  The liquid volume can be increase by injecting the liquid into the droplet.  (b) A volatile droplet (critical nucleus of heterogeneous nucleation) on a flat and smooth substrate.  Now, the liquid volume can be changed indirectly by controlling the pressure of the surrounding vapor.  
}
\label{fig:L3}
\end{center}
\end{figure}

Consider the Helmholtz free energy of a spherical cap-shaped droplet of a nonvolatile liquid with the radius $R$ and the contact angle $\theta$ on a flat smooth substrate (Fig.~\ref{fig:L1}) given by
\begin{eqnarray}
F &=& S_{\rm LV}\sigma + S_{\rm SL}\Delta\sigma+L_{\rm SLV}\tau 
\label{eq:T4}
\end{eqnarray}
where $S_{\rm LV}$, $S_{\rm SL}$, and $L_{\rm SLV}$ defined by
\begin{eqnarray}
S_{\rm LV} &=& 2\pi R^{2}\left(1-\cos\theta\right)
\label{eq:T5} \\
S_{\rm SL} &=& \pi R^{2}\sin^{2}\theta
\label{eq:T6} \\
L_{\rm SLV} &=& 2\pi R\sin\theta
\label{eq:T7}
\end{eqnarray}
are the liquid-vapor surface area, the liquid-substrate surface area, and the substrate-liquid-vapor three phase contact line length respectively.  The liquid-vapor surface tension is denoted by $\sigma$,  and the intrinsic line tension is denoted by $\tau$.  Since, we consider a nonvolatile liquid droplet, we use Helmholtz free energy instead of the Gibbs or grand potential free energy~\cite{Kondo1955,Rusanov2004,Schimmele2007,Wang2014}.  Therefore, the problem of dividing surface of liquid-vapor interface~\cite{Rowlinson1982,Rusanov2004,Schimmele2007} and the Kondo equation~\cite{Kondo1955,Rusanov2004} for the Laplace pressure will not be considered explicitly.  Here, the droplet volume is controlled by injecting or subtracting liquid from the nonvolatile droplet (Fig.~\ref{fig:L3}(a)).  By contrast, the volatile droplet is realized as a critical nucleus of the heterogeneous nucleation so that the droplet volume is controlled indirectly only when the supersaturation of surrounding vapor is altered (Fig.~\ref{fig:L3}(b)).  Also, the volatile droplet is not stable but is a metastable transient state of nucleation process.

The difference between the substrate-liquid (SL) surface tension $\sigma_{\rm SL}$ and the substrate-vapor (SV) surface tensions $\sigma_{\rm SV}$ is given by $\Delta\sigma$:
\begin{equation}
\Delta\sigma = \sigma_{\rm SL}-\sigma_{\rm SV}.
\label{eq:T8}
\end{equation}
In the capillary approximation, the liquid-vapor surface tension $\sigma$ is assumed to be affected neither by the presence of the substrate~\cite{MacDowell2013} nor the curvature of the meniscus~\cite{Tolman1949,Iwamatsu1994,vanGeesen1998,Liu2013,Kanduc2017}.  Similarly, the liquid-substrate surface tension $\sigma_{\rm SL}$ corresponds to the interaction energy between the substrate and the infinitely thick wetting layer.  Also, the line tension $\tau$ is assumed to be a constant.

By minimizing the Helmholtz free energy in Eq.~(\ref{eq:T4}) under the subsidiary condition of constant volume
\begin{equation}
V=\frac{4}{3}\pi R^{3} \phi\left(\theta\right)
\label{eq:T9}
\end{equation}
with
\begin{equation}
\phi\left(\theta\right)=\frac{\left(2+\cos\theta\right)\left(1-\cos\theta\right)^{2}}{4},
\label{eq:T10}
\end{equation}
we can derive the modified Young's equation (\ref{eq:T1}).

To this end, we use the formula~\cite{Shapiro2003}
\begin{equation}
\frac{dF}{d\theta}=\left(\frac{\partial F}{\partial R}\right)Rq\left(\theta\right)+\left(\frac{\partial F}{\partial \theta}\right)
\label{eq:T11}
\end{equation}
with the geometrical factor $q\left(\theta\right)$ defined by
\begin{equation}
q\left(\theta\right)=-\frac{\left(1+\cos\theta\right)\sin\theta}{\left(1-\cos\theta\right)\left(2+\cos\theta\right)},
\label{eq:T12}
\end{equation}
and we obtain the equilibrium condition
\begin{equation}
\Delta\sigma+\sigma\cos\theta+\frac{\tau}{R\sin\theta}=0.
\label{eq:T13}
\end{equation}
from
\begin{equation}
\left(-\frac{2+\cos\theta}{2\pi R^{2}\sin\theta}\right)\frac{dF}{d\theta}=0.
\label{eq:T14}
\end{equation}

Using Young's equation with Young's contact angle $\theta_{\rm Y}$ given by
\begin{equation}
\Delta\sigma= \sigma_{\rm SL}-\sigma_{\rm SV}=-\sigma\cos\theta_{\rm Y},
\label{eq:T15}
\end{equation}
which expresses the force-balance at the three-phase contact line without the effect of line tension, Eq.~(\ref{eq:T13}) is written as
\begin{equation}
\cos\theta = \cos\theta_{\rm Y}-\frac{\tau}{\sigma R\sin\theta},
\label{eq:T16}
\end{equation}
which is equivalent to Eq.~(\ref{eq:T1}) since $r=R\sin\theta$ is a contact line radius (Fig.~\ref{fig:L1}).

If the liquid-vapor surface tension $\sigma$ depends on the radius $R$, and the intrinsic line tension $\tau$ depends on the  contact angle $\theta$, the equilibrium condition from the minimization of the free energy in Eq.~(\ref{eq:T14}) gives 
\begin{equation}
\Delta\sigma + \sigma\cos\theta + \frac{\tau}{R\sin\theta}
 + \left(1+\cos\theta\right)R\left(\frac{\partial\sigma}{\partial R}\right)
 -\frac{2+\cos\theta}{R}\left(\frac{\partial\tau}{\partial \theta}\right)=0
\label{eq:T17}
\end{equation}
instead of Eq.~(\ref{eq:T13}), where we have used partial derivative ($\partial$) to emphasize that the surface tension $\sigma$ and the line tension $\tau$ can depend on various parameters in addition to $R$ and $\theta$.  Note that those two derivatives $\partial\sigma/\partial R$ and $\partial\tau/\partial \theta$ do not come from any expansion of the original modified Young's equation in Eq.~(\ref{eq:T1}), but they come directly from the free energy minimization.  Therefore, only the fist derivatives appear.  Also, the derivative $\partial\sigma/\partial R$ is the change of the liquid-vapor surface tension by the {\it real} displacement of the {\it real} surface, and does not corresponds to $\left[\partial\sigma/\partial R\right]$ of the {\it fictitious} displacement of the dividing surface~\cite{Kondo1955,Rowlinson1982,Rusanov2004,Schimmele2007}.

Suppose the liquid-vapor surface tension $\sigma$ depends on the radius $R$ through Tolman's formula
\begin{equation}
\sigma=\sigma_{\infty}\left(1-\frac{2\delta_{\rm T}}{R}\right)
\label{eq:T18}
\end{equation}
where $\delta_{\rm T}$ is Toman's length.  Then, Young's equation in Eq.~(\ref{eq:T15}) should be replaced by
\begin{equation}
\Delta\sigma= \sigma_{\rm SL}-\sigma_{\rm SV}=-\sigma_{\infty}\cos\theta_{\rm Y},
\label{eq:T19}
\end{equation}
and we can transform Eq.~(\ref{eq:T17}) into the form of the modified Young's equation in Eq.~(\ref{eq:T1}): 
\begin{equation}
\cos\theta = \cos\theta_{\rm Y}-\frac{\tau_{\rm app}}{\sigma_{\infty} r},
\label{eq:T20}
\end{equation}
where the {\it apparent} line tension $\tau_{\rm app}$ is given by
\begin{equation}
\tau_{\rm app} = \tau  + 2\sin\theta\delta_{\rm T}\sigma_{\infty} 
- \left(2+\cos\theta\right)\sin\theta\left(\frac{\partial\tau}{\partial\theta}\right).
\label{eq:T21}
\end{equation}
Therefore, if the correction $2\delta_{\rm T}\sigma_{\infty}$ and $\partial\tau/\partial\theta$ are the same order of magnitude as the intrinsic line tension $\tau$, the measured apparent line tension $\tau_{\rm app}$ from the modified Young's equation (\ref{eq:T20}) does not represent the intrinsic line tension $\tau$ unless the contact angle $\theta$ approaches zero ($\theta\rightarrow 0$).   A similar expression was derived by Marmur~\cite{Marmur1997} though he used the ratio
\begin{equation}
\alpha=\frac{1-\cos\theta}{\sin\theta}
\label{eq:T22}
\end{equation}
instead of the contact angle $\theta$.  His result (Eq. [23] in reference~\cite{Marmur1997})  reduces to the third term of the right-hand side of Eq.~(\ref{eq:T21}). 

An expression similar to Eq.~(\ref{eq:T21}) for two-dimensional (2D) cylindrical droplets is derived by Kandu\v{c}~\cite{Kanduc2017} to discuss the effect of the curvature dependence of the liquid-vapor surface tension $\partial\sigma/\partial R$ and the contact-angle dependence of the intrinsic line tension $\partial\tau/\partial\theta$ on the apparent line tension $\tau_{\rm app}$.   Using the curvature-dependent liquid-vapor surface tension from the molecular dynamics (MD) simulation, Kandu\v{c} deduced the magnitude of Tolman's length $\delta_{\rm T}\simeq -0.05$nm for the model water.  Tolaman's length with the same negative sign and a similar magnitude have bee suggested for three-dimensional (3D) droplets by mean-field calculations based on the density functional theory~\cite{Iwamatsu1994,vanGeesen1998} and by similar MD simulations~\cite{Block2010,Khalkhali2017}.  However, the radius-dependence of the Tolman's length and, therefore, the higher order $1/R^{2}$ correction to the liquid-vapor surface tension is appreciable~\cite{Block2010}.  Also, it is difficult to determine Tolman's length experimentally from nucleation data~\cite{Bruot2016}. Together with the liquid-vapor surface tension $\sigma\sim 0.06$ N/m of water droplets~\cite{Kanduc2017,Khalkhali2017}, for example, the correction $2\sin\theta\delta_{\rm T}\sigma_{\infty}$ amounts to $\sim \left(-6\times 10^{-12} {\rm N}\right)\sin\theta$ from Eq.~(\ref{eq:T21}) for spherical cap-shaped droplets, which is the same order of magnitude as the atomic $\tau_{\rm atom}$ and the van der Waals $\tau_{\rm vdW}$ line tensions.  The sign of this correction is determined from that of Tolman's length $\delta_{\rm T}$.

By contrast, the study of contact-angle dependence of the intrinsic line tension is scarce~\cite{Marmur1997,Heim2013,Kanduc2017}.   Very recently, Kand\v{u}c~\cite{Kanduc2017} has conducted the MD simulation of cylindrical droplet and deduced the derivative $\partial\tau/\partial\theta$.  This derivative itself shows strong contact-angle ($\theta$) dependence and is on the order of $\sim -15\times 10^{-12}$ N, which is the same order of magnitude as the line tensions $\tau_{\rm atom}$ and $\tau_{\rm vdW}$.  This derivative is large negative for hydrophilic substrates ($\cos\theta>0$) and is small positive for hydrophobic substrates ($\cos\theta<0$)~\cite{Kanduc2017}.

Although equation (\ref{eq:T17}) takes into account the curvature dependence of the liquid-vapor surface tension $\sigma=\sigma\left(R\right)$ and the contact angle dependence of the intrinsic line tension $\tau=\tau\left(\theta\right)$, the effect of the substrate potential or other external field is not explicitly taken into account.  In fact, equation (\ref{eq:T4}) can be written as
\begin{equation}
F = 2\pi R^{2}\left(1-\cos\theta\right)\sigma+\pi R^{2}\sin^{2}\theta \Delta\sigma + \Delta F\left(R, \theta\right),
\label{eq:T23}
\end{equation}
where $\Delta F\left(R, \theta\right)$ stands for the excess free-energy, which takes into account various effects such as the vapor-liquid-substrate interaction and external fields.  Then, we obtain the generalization of Young's equation
\begin{equation}
\Delta\sigma + \sigma\cos\theta
 + \left(1+\cos\theta\right)R\left(\frac{\partial\sigma}{\partial R}\right)_{\theta}
+\left(-\frac{2+\cos\theta}{2\pi R^{2}\sin\theta}\right)\left(\left(\frac{\partial \Delta F}{\partial R}\right)_{\theta}Rq\left(\theta\right)+\left(\frac{\partial \Delta F}{\partial \theta}\right)_{R}\right)=0 
\label{eq:T24}
\end{equation}
or
\begin{equation}
\sigma\cos\theta = \sigma_{\infty}\cos\theta_{\rm Y}
 - \left(1+\cos\theta\right)R\left(\frac{\partial\sigma}{\partial R}\right)_{\theta}
-\left(-\frac{2+\cos\theta}{2\pi R^{2}\sin\theta}\right)\left(\left(\frac{\partial \Delta F}{\partial R}\right)_{\theta}Rq\left(\theta\right)+\left(\frac{\partial \Delta F}{\partial \theta}\right)_{R}\right)
\label{eq:T25}
\end{equation}
instead of Eq.~(\ref{eq:T17}) from Eqs.~(\ref{eq:T11}), (\ref{eq:T12}), (\ref{eq:T14}), and (\ref{eq:T19}).  Equation (\ref{eq:T25}) is the most general form of a generalized Young's equation.  We can recover Eq.~(\ref{eq:T17}) when 
\begin{equation}
\Delta F\left(R,\theta\right)=2\pi R\sin\theta\tau\left(\theta\right).
\label{eq:T26}
\end{equation}  
Most of previous microscopic  calculations of line tension assumed this form and concentrated on calculating this intrinsic line tension $\tau\left(\theta\right)$~\cite{Indekeu1992,Getta1998,Pompe2002,Herminghaus2006}. 

Equation~(\ref{eq:T25}) can be transformed into the form of Eq.~(\ref{eq:T20}), and the apparent line tension is given formally by
\begin{equation}
\tau_{\rm app} =  2\sin\theta\delta_{\rm T}\sigma_{\infty}
+ \left(-\frac{2+\cos\theta}{2\pi R}\right)\left(\left(\frac{\partial \Delta F}{\partial R}\right)_{\theta}Rq\left(\theta\right)+\left(\frac{\partial \Delta F}{\partial \theta}\right)_{R}\right),
\label{eq:T27}
\end{equation}
where Tolman's formula (\ref{eq:T18}) is used, which reduces to Eq.~(\ref{eq:T21}) when Eq.~(\ref{eq:T26}) is used.  Note that Eq.~(\ref{eq:T27}) does not necessarily gives the definition of  line tension because it may not give the $1/r$ correction to the contact angle in Eq.~(\ref{eq:T1}).

\section{van der Waals and gravitational line tensions}

\subsection{van der Waals line tension}

When the droplet is nanometer-scale, the spherical cap-shape approximation will be valid and the model developed in the previous section will be applicable.  Marmur~\cite{Marmur1997} derived the substrate-liquid-vapor (vacuum) interaction energy $U_{\rm SLV}$
\begin{equation}
U_{\rm SLV} =-\frac{A_{\rm SL}}{6}\left[\frac{R^2-y_{0}^{2}}{2\delta^{2}} +2\frac{y_{0}}{\delta} + \ln\frac{\delta_{\rm vdW}}{R+y_{0}} 
- \frac{R+3y_{0}}{2\left(R+y_{0}\right)}\right]
\label{eq:T28}
\end{equation}
between the substrate and the spherical cap-shaped droplet directly by assuming the van der Waals type attractive interaction, where $y_{0}=-R\cos\theta$ (Fig.~\ref{fig:L1}) is the vertical position of the center of a spherical droplet, $A_{\rm SL}$ represents the Hamaker constant and $\delta$ represents the cut-off distance of the long-ranged van der Waals potential~\cite{Marmur1997,Israelachvili2011}.  Then, the solid-liquid surface tension is written as~\cite{Marmur1997,Israelachvili2011}
\begin{equation}
\sigma_{\rm SL}=\frac{A_{\rm SL}}{24\pi\delta^{2}}.
\label{eq:T29}
\end{equation}
and the excess free energy $\Delta F$ in Eq.~(\ref{eq:T23}) is given by
\begin{eqnarray}
\Delta F\left(R,\theta\right) &=& U_{\rm SLV}-S_{\rm SL}\sigma_{\rm SL}
\nonumber \\
&=& -\frac{A_{\rm SL}}{6}\left[2\frac{y_{0}}{\delta} + \ln\frac{\delta}{R+y_{0}} - \frac{R+3y_{0}}{2\left(R+y_{0}\right)}\right],
\nonumber \\
\label{eq:T30}
\end{eqnarray}
where we have used the fact that the liquid-substrate surface area $S_{\rm SL}$ in Eq.~(\ref{eq:T6}) is also written as $S_{\rm SL}=\pi\left(R^{2}-y_{0}^{2}\right)$ (Fig.~\ref{fig:L1}).

Marmur~\cite{Marmur1997} identified $\tau=\Delta F\left(R,\theta\right)/L_{\rm SLV}$ given by
\begin{eqnarray}
\tau &=& \frac{A_{\rm SL}}{12\pi R \sin\theta}\left(-2\frac{y_{0}}{\delta} 
-\ln\frac{\delta}{R+y_{0}} 
+ \frac{R+3y_{0}}{2\left(R+y_{0}\right)} \right) \nonumber \\
&\approx& \frac{A_{\rm SL}}{6\pi\delta}\cot\theta
\label{eq:T31}
\end{eqnarray}
as the intrinsic van der Waals line tension, where only the first term is retained and $y_{0}=-R\cos\theta$ (Fig.~\ref{fig:L1}) is used.  Since this intrinsic line tension in Eq.~(\ref{eq:T31}) depends only on the contact angle $\theta$ and the excess free energy is approximately written as Eq.~(\ref{eq:T26}), it is possible to calculate the correction to the apparent line tension $\tau_{\rm app}$ from Eq.~(\ref{eq:T21}) using
\begin{equation}
\frac{\partial \tau}{\partial \theta}=-\frac{A_{\rm SL}}{6\pi\delta\sin^{2}\theta}.
\label{eq:T32}
\end{equation}
Then, the apparent van der Waals line tension becomes
\begin{equation}
\tau_{\rm vdW} \approx \frac{A_{\rm SL}\sin\theta}{3\pi\delta\left(1-\cos\theta\right)}.
\label{eq:T33}
\end{equation}
from Eq.~(\ref{eq:T21}) where we have neglected $\partial \sigma/\partial R$. 

In fact, the correct formula for the apparent van der Waals line tension is given exactly by
\begin{eqnarray}
\tau_{\rm vdW} &=& \frac{A_{\rm SL}}{12\pi}\left[\frac{4\sin\theta}{\delta\left(1-\cos\theta\right)} - \frac{3\sin\theta}{R\left(1-\cos\theta\right)^{2}} \right]
\nonumber \\
&\approx& \frac{A_{\rm SL}\sin\theta}{3\pi\delta\left(1-\cos\theta\right)}
\label{eq:T34}
\end{eqnarray}
from Eq.~(\ref{eq:T27}), where we have neglected $\partial \sigma/\partial R$ again.  Note that most of the singular terms in Eq.~(\ref{eq:T31}) disappears automatically in Eq.~(\ref{eq:T34}).

Although the intrinsic line tension $\tau$ in Eq.~(\ref{eq:T31}) becomes negative for hydrophobic droplets with $\theta>90^{\circ}$, the apparent line tension in Eqs.~(\ref{eq:T33}) and (\ref{eq:T34}) is always positive.  The derivative $\partial\tau/\partial\theta$ is the same order of magnitude as the intrinsic line tension $\tau$, and make the apparent line tension $\tau_{\rm app}$ always positive.  This result is consistent to the MD results of Kandu\v{c}~\cite{Kanduc2017} for a cylindrical droplet on a hydrophilic substrate. Of course, the apparent line tension $\tau_{\rm app}$ also depends strongly on the contact angle $\theta$ as expected~\cite{Heim2013}. In particular, the apparent line tension diverges as $\tau_{\rm app}\rightarrow 2A_{\rm SL}/3\pi\delta \theta \sim 1/\theta\rightarrow \infty$ in the limit of complete wetting $\theta\rightarrow 0$.

Since the direct estimation of the solid-liquid surface tension $\sigma_{\rm SL}$ or the Hamaker constant $A_{\rm SL}$ is difficult, it is customary to use the combining relation~\cite{Marmur1997,Israelachvili2011} to simplify Eq.~(\ref{eq:T29}).  However, we will not modify~\cite{Marmur1997} Eqs.~(\ref{eq:T24}) and (\ref{eq:T29}) further since we are mainly interested in the contact-angle dependence of line tension.  We merely note that $\tau_{\rm vdW}\sim \sigma_{\rm SL}\delta$.  Therefore,  $\left|\tau_{\rm vdW}\right|\sim 1-100$pN as $\delta\sim 0.165$nm~\cite{Israelachvili2011}.   Although the apparent van der Waals line tension in Eq.~(\ref{eq:T34}) is always positive, the total line tension defined in Eqs.~(\ref{eq:T21}) or (\ref{eq:T27}) can be either positive or negative as Tolman's length $\delta_{\rm T}$ can be negative~\cite{Iwamatsu1994,vanGeesen1998,Block2010,Khalkhali2017}.

\subsection{Gravitational line tension}

When the droplet is large and is of the size of capillary length $\kappa^{-1}=\sqrt{\sigma_{\infty}/\rho g}\sim 1$ mm, where $\rho$ is the density of the liquid and $g$ is the acceleration of gravity, the gravitational potential exceeds the van der Waals potential.  Then, the line tension is dominated by the gravitational line tension $\tau_{\rm grav}$ (Eq.~(\ref{eq:T3})).  Using the spherical cap-shaped droplet model, the excess free energy is given by~\cite{Shapiro2003}
\begin{equation}
\Delta F = \frac{\pi R^{4}\rho g}{12}\left(3+\cos\theta\right)\left(1-\cos\theta\right)^{3}
\label{eq:T35}
\end{equation}
Then Eq.~(\ref{eq:T27}) gives
\begin{equation}
\tau_{\rm app}=-\frac{R^{3}\rho g}{6}\left(1-\cos\theta\right)^{2}\sin\theta,
\label{eq:T36}
\end{equation}
which is always negative, where we have neglected the curvature dependence of the liquid-vapor surface tension $\partial\sigma/\partial R$ because it is characterized by Tolman's length and is expected to be much smaller than Eq.~(\ref{eq:T36}).  The negativity of Eq.~(\ref{eq:T36}) is intuitively correct because the droplet will spreads due to the gravitational acceleration  and the contact angle will become lower.  For fixed droplet volume $V$ given by Eq.~(\ref{eq:T9}), Eq.~(\ref{eq:T36}) can also be written as
\begin{equation}
\tau_{\rm app}=-\frac{M g \sin\theta}{2\pi\left(2+\cos\theta\right)\left(1-\cos\theta\right)}
\label{eq:T37}
\end{equation}
where $M=V\rho$ is the mass of the droplet.  In contrast to the van der Waals line tension, Eq.~(\ref{eq:T37}) is always negative.  It diverges as $\tau_{\rm app}\rightarrow -Mg/2\pi\theta \sim -1/\theta$ in the limit of complete wetting $\theta\rightarrow 0$.  When $\theta=90^{\circ}$, for example, Eq.~(\ref{eq:T37}) becomes $\tau_{\rm app}=-Mg/4\pi$ whose magnitude is $\tau_{\rm app}\sim 10^{-5}{\rm N}$ for $V=10 {\rm mm}^{3}$ water droplet.  This is the right order of magnitude of line tension usually observed for macroscopic millimeter-scale droplets.~\cite{Heim2013,David2007,Law2017}.  Therefore, the curvature dependence of the liquid-vapor surface tension represented by Tolman's length, will be unimportant for the gravitational line tension and the correction due to $\partial\sigma/\partial R$ or $\delta_{\rm T}$ can be neglected.

However, the results in Eqs.~(\ref{eq:T36}) and (\ref{eq:T37}) cannot be interpreted as the gravitational line tension $\tau_{\rm grav}$ because $\tau_{\rm app}\propto R^{3}$ which does not give the $1/r$ correction to $\cos\theta$ in Eq.~(\ref{eq:T1}). In fact, Eq.~(\ref{eq:T25}) becomes~\cite{Shapiro2003}
\begin{equation}
\cos\theta=\cos\theta_{\rm Y}+\frac{R^{2}\rho g}{6\sigma}\left(1-\cos\theta\right)^{2}
\label{eq:T38}
\end{equation}
from Eqs.~(\ref{eq:T15}), (\ref{eq:T24}) and (\ref{eq:T35}), which does not have the form of Eq.~(\ref{eq:T1}) because $R\propto r$.  Eq.~(\ref{eq:T38}) is further simplified if the contact angle is small ($\theta\ll 1$ and $\theta_{\rm Y}\ll 1$).  Then, we have
\begin{equation}
\theta \simeq \sqrt{1-\frac{R^{2}\rho g}{6\sigma}}\theta_{\rm Y},
\label{eq:T39}
\end{equation}
which predicts $\theta\simeq\sqrt{1-0.25}\theta_{\rm Y}\simeq 0.87\theta_{\rm Y}$ for a $R\sim 3$mm water droplet with $\sigma\sim 0.06$N/m.  Therefore, the droplet will spread and the contact angle decreases by the action of the gravity, which could be interpreted as the effect of negative gravitational line tension from Eq.~(\ref{eq:T1})

So far we have assumed a spherical cap-shaped droplet.  When the droplet becomes larger, the droplet will be flattened and it becomes paddle-shaped~\cite{deGennes2004} and the spherical cap-shaped model to derive Eq.~(\ref{eq:T38}) from Eq~(\ref{eq:T35}) is no longer valid.  In such a case, it is customary to assume the excess free energy in the form of Eq.~(\ref{eq:T26}) and to calculate the intrinsic line tension $\tau\left(\theta\right)$ using the two-dimensional wedge-like model with varying sophistications~\cite{Pompe2000,Schimmele2007,Indekeu1992,Getta1998,Pompe2002}

\begin{figure}[htbp]
%Fig.3
\begin{center}
\includegraphics[width=0.85\linewidth]{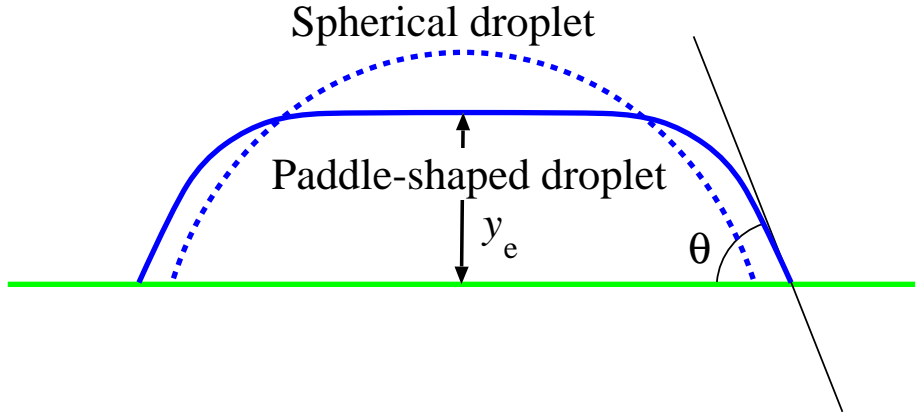}
\caption{
A millimeter-scale paddle-shaped droplet under the gravitational acceleration.  The spherical meniscus of nanometer-scale droplet in Fig.~\ref{fig:L1} is distorted by the gravitational force.
}
\label{fig:L2}
\end{center}
\end{figure}

The gravitational intrinsic line tension for a paddle-shaped droplet (Fig.~\ref{fig:L2}) has already been obtained using the interface displacement model~\cite{Herminghaus2006,deGennes2004}.  The intrinsic line tension $\tau\left(\theta\right)$ is given as
\begin{equation}
\tau\left(\theta\right)=\frac{1}{2}\sigma_{\infty} y_{\rm e}\theta=\frac{1}{2}\sigma_{\infty}\kappa^{-1}\theta^{2},
\label{eq:T40}
\end{equation}
when the contact angle $\theta$ is small, where 
\begin{equation}
y_{\rm e}=2\kappa^{-1}\sin\left(\frac{\theta}{2}\right)\simeq \kappa^{-1}\theta
\label{eq:T41}
\end{equation}
is the height (thickness) of the paddle-shaped droplet~\cite{Andrieu1994,deGennes2004} shown in Fig.~\ref{fig:L2}.   Apparently, the intrinsic line tension $\tau\left(\theta\right)$ is positive and would be several order of magnitude large than the van der Waals line tension $\tau_{\rm vdW}$ as the capillary length $\kappa^{-1}\sim 1 {\rm mm}$ is several order of magnitude larger~\cite{Law2017} than the cut-off distance~\cite{Israelachvili2011} $\delta\sim 0.165 {\rm nm}$ (see the discussion below Eq.~(\ref{eq:T29})).  A slightly more general, but essentially the same result for $\tau\left(\theta\right)$ is derived by Law {\it et al}~\cite{Law2017}.  

This intrinsic line tension $\tau\left(\theta\right)$ in Eq.~(\ref{eq:T41}) successfully accounts for the large line tensions of the order of $\mu{\rm N}$ observed for millimeter-scale large droplets~\cite{Law2017,David2007}.  However, this intrinsic line tension is always positive, which contradicts to the negative effective line tension in Eq.~(\ref{eq:T37}).  Also, the experimentally determined line tensions are frequently negative~\cite{David2007}. 

In fact, the apparent gravitational line tension must contain the correction from $\partial \tau/\partial\theta$, and it becomes
\begin{eqnarray}
\tau_{\rm grav} &=& \tau  - \left(2+\cos\theta\right)\sin\theta\left(\frac{\partial\tau}{\partial\theta}\right).
\nonumber \\
&=& \left(1-\frac{2\left(2+\cos\theta\right)\sin\theta}{\theta}\right)\tau \simeq -5\tau
\label{eq:T42}
\end{eqnarray}
from Eq.~(\ref{eq:T21}). Now, the apparent gravitational line tension $\tau_{\rm grav}$ becomes always negative because the intrinsic line tension $\tau$ in Eq.~(\ref{eq:T40}) is always positive.  This result is consistent to the result in Eq.~(\ref{eq:T37}). 

The sign of experimentally measured line tensions are either positive or negative~\cite{Drelich1996,David2007}.  Some group~\cite{Drelich1993} reported negative line tensions of the order of $\mu {\rm N}$, while other group~\cite{Amirfazli2000} reported positive line tensions of the same order of magnitude for millimeter-scale droplets.   However, most measurements of negative line tension are for large millimeter-scale droplets~\cite{David2007}.  The ambiguity regarding the sign and the magnitude of line tension is known to be partly caused by the roughness and heterogeneity of the real surface~\cite{Law2017}.  Actually, simple pinning mechanism will easily give a spurious positive line tension.  However, the negative line tension cannot stabilize the three-phase contact line so that the higher-order correction must necessarily be important~\cite{Berg2010}.

We have been considering the problem of droplet using the Helmholtz free energy of fixed volume.   Therefore, the droplet volume is changed by injecting liquid into or subtracting liquid from the droplet as shown in Fig.~\ref{fig:L3}(a).  Then, the Helmholtz free energy is minimized to seek for the equilibrium contact angle $\theta$ for fixed volume.  Very recently, Tadmor et al.~\cite{Tadmor2018} have discussed the gravitational line tension by using the variation of the Gibbs free energy, which allows for the fluctuation of droplet volume.  They derived two formulas for the cosine of the contact angle $\theta$.  The one formula derived from the variation of Gibbs free energy by the contact angle $\theta$ under the condition of fixed volume coincides with Eq.~(\ref{eq:T38}).  In addition, they derived another new formula from the variation of the free energy by the volume under the condition of (arbitrary) fixed contact angle.  They claimed that these two solutions suggest two coexisting stable contact angles.  However, the new formula which determines the contact angle $\theta$ is mathematically contradictory as the variation is performed under the condition of arbitrary fixed contact angle $\theta$.

In order to consider the variation by the contact angle as well as by the volume or the number of molecule in the droplet simultaneously, it is necessary to regard the droplet as the critical nucleus of heterogeneous nucleation~\cite{Rusanov2004}, which is in metastable equilibrium with the oversaturated surrounding vapor (Fig.~\ref{fig:L3}(b)).  In this case, the droplet volume is controlled indirectly by changing the pressure or the supersaturation of surrounding vapor (Fig.~\ref{fig:L3}(b)).  Tatyanenko and Shchekin~\cite{Tatyanenko2017} studied the size-dependence of the contact angle of this critical nucleus recently.  By extremizing the grand potential instead of the Helmholtz free energy by the volume and by the contact angle, they derived the Laplace formula for the pressure as well as the generalized Young's equation for the contact angle $\theta$.  They~\cite{Tatyanenko2017} further used the microscopic interface-displacement model to study the size dependence of the contact angle of critical nucleus.  They found a large deviation from the linearity $\cos\theta\propto 1/r$.  The size dependence of the contact angle is attributed not only to the line tension but to the adsorption of wetting layer (Fig.~\ref{fig:L3}(b)) since the adsorption depends on the supersaturation.  Although their result is suggestive to some of experimental results~\cite{Checco2003}, it does not include Tolman's correction and does not provide any general conclusion since their numerical result depends strongly on the choice of disjoining pressure.

In this paper, we have been considering a sessile droplet on flat and ideally smooth substrates.  A general equation similar to the modified Young's equation that can predicts the apparent contact angle $\theta$ on a flat rough substrate including the line tension effect was proposed by Bormashenko~\cite{Bormashenko2011}.  However, the result of recent molecular dynamic simulation~\cite{Wloc2018} indicates negligible contribution of line tension.  The various effects of roughness, surface imperfections and impurities, which would obscure the line tension effect by the contact angle hysteresis, are beyond the scope of this paper.

\section{\label{sec:sec5}Conclusion}

In this study, we derived a generalized Young's equation which includes the two corrections due to the curvature-dependence of the liquid-vapor surface tension and the contact-angle dependence of line tension.  We considered the curvature-dependence using Tolman's formula~\cite{Tolman1949,Iwamatsu1994}.  We considered the contact-angle dependence of the van der Waals line tension in nanometer scale by employing the simple model by Marmur~\cite{Marmur1997} which takes into account the long-ranged van der Waals potential.  For the gravitational line tension of larger millimeter-scale droplets, we used the formula for the excess free energy developed by Shapiro et al.~\cite{Shapiro2003} and that for the intrinsic line tension developed by de Gennes et al.~\cite{deGennes2004} and by Herminghaus and Brochard~\cite{Herminghaus2006}.  

In conclusion, our analysis shows that the two corrections to the apparent line tension are large and cannot be neglected.  They change even the sign of the resultant apparent line tension.  Qualitatively, our two simple model calculations explain the variety of the sign and the magnitude of line tensions experimentally measured~\cite{David2007}. If the line tension can be artificially controlled, it can be used to change the contact angle and the wetting properties of the substrate.  In fact, a modest seasonal change of line tension leads to the seasonal change of the cuticle wettability of insects from superhydrophobic to superhydrophilic (wetting)~\cite{Gundersen2017}.  Our theoretical consideration of line tension will be useful in future to design substrates whose wetting property is controlled by line tension.  

Finally, we would like to emphasize again that the theoretical intrinsic line tension is defined through the free energy~\cite{Schimmele2007,Schimmele2009,Indekeu1992,Getta1998,Pompe2002,Herminghaus2006}, while the experimental apparent line tension is determined from the generalized Young's equation of the mechanical force balance~\cite{Law2017,Drelich1996,Pompe2000,Wang2001,Checco2003,Kameda2008,Berg2010}.   In other word, there is a gap between the theoretically line tensions defined through the {\it energy} and the experimentally line tensions define trough the {\it force}.  Therefore,  naive comparison of the experimentally determined and the theoretically calculated line tension is not always possible.

\section*{Acknowledgements}

A part of this work was conducted as a visiting scientist at the Department of Physics, Tokyo Metropolitan University.  The author is grateful to Prof. Hiroyuki Mori and Prof. Yutaka Okabe (Tokyo Metropolitan Univ.) for continuous support and encouragement.  The author is also grateful to Dr. Lothar Schimmele and Prof. Siegfried Dietrich (Max-Planck Institute for Intelligent Systems, Stuttgart) for their useful comments on the initial version of the manuscript.  He is also grateful to Prof. Dmitry V. Tatyanenko and Prof. Alexander K. Shchekin (St. Petersburg State Univ.) for useful discussion about their new result in reference [48].  He also thanks to the two reviewers for their useful suggestions.

\end{document}